\documentclass[3p,twocolumn]{elsarticle}

\usepackage{lineno,hyperref}
\modulolinenumbers[5]

\journal{Nucl. Instru. Methods Phys. Res. A}

\begin{document}

\begin{frontmatter}

\title{Performance recovery of long CsI(Tl) scintillator crystals with APD-based readout}

\author[myaddress,mysecondaryaddress]{P. Cabanelas \corref{correspondingauthor}}
\cortext[correspondingauthor]{Corresponding author}
\ead{pablo.cabanelas@usc.es}

\author[myaddress,mysecondaryaddress]{D. Gonz\'alez}
\author[myaddress,mysecondaryaddress]{H. Alvarez-Pol}
\author[myaddress,mysecondaryaddress]{J.M. Boillos}
\author[uvigo]{E. Casarejos}
\author[lund]{J. Cederkall}
\author[myaddress,mysecondaryaddress]{D. Cortina}
\author[myaddress,mysecondaryaddress]{M. Feijoo}
\author[lip,fcul]{D. Galaviz}
\author[myaddress,lip]{E. Galiana}
\author[tum]{R. Gernh\"auser}
\author[lund]{P. Golubev}
\author[tuda]{A.-L. Hartig}
\author[tum]{P. Klenze}
\author[lund]{A. Knyazev}
\author[tuda]{T. Kr\"oll}
\author[ific]{E. N\'acher}
\author[lund]{J. Park}
\author[csic]{A. Perea}
\author[myaddress,mysecondaryaddress]{B. Pietras}
\author[tum]{L. Ponnath}
\author[tuda]{H.-B. Rhee}
\author[myaddress,mysecondaryaddress]{J.L. Rodr\'iguez-S\'anchez}
\author[tuda]{C. Suerder}
\author[csic]{O. Tengblad}
\author[lip,fcul]{P. Teubig}

\address[myaddress]{Instituto Galego de F\'isica de Altas Enerx\'ias, University of Santiago de Compostela, E-15782 Santiago de Compostela, Spain}
\address[mysecondaryaddress]{Particle Physics Department, University of Santiago de Compostela, E-15782 Santiago de Compostela, Spain}
\address[uvigo]{Universidade de Vigo, E-36310 Vigo, Spain}
\address[tuda]{Technische Universit\"at Darmstadt, Fachbereich Physik, Institut f\"ur Kernphysik, D-64289 Darmstadt, Germany}
\address[lip]{Laboratory for Instrumentation and Experimental Particle Physics, LIP, 1649-003 Lisbon, Portugal}
\address[fcul]{Departamento de F\'isica, Faculdade de Ci\^encias da Universidade de Lisboa, 1749-016 Lisbon, Portugal}
\address[lund]{Department of Physics, Lund University, SE-221 00 Lund, Sweden}
\address[tum]{Physik Department, Technische Universit\"at M\"unchen, 85748 Garching, Germany}
\address[csic]{Instituto de Estructura de la Materia, CSIC, E-28006 Madrid, Spain}
\address[ific]{Instituto de F\'isica Corpuscular, CSIC - Universitat de Valencia, E-46980, Valencia, Spain}

\begin{abstract}
CALIFA is the high efficiency and energy resolution calorimeter for the R$^{3}$B experiment at FAIR, intended for detecting high energy light charged particles and gamma rays in scattering experiments, and is being commissioned during the Phase-0 experiments at FAIR, between 2018 and 2020. It surrounds the reaction target in a segmented configuration with 2432 detection units made of long CsI(Tl) finger-shaped scintillator crystals. CALIFA has a 10 year intended operational lifetime as the R$^{3}$B calorimeter, necessitating measures to be taken to ensure enduring performance. In this paper we present a systematic study of two groups of 6 different detection units of the CALIFA detector after more than four years of operation. The energy resolution and light output yield are evaluated under different conditions. Tests cover the aging of the first detector units assembled and investigates recovery procedures for degraded detection units. A possible reason for the observed degradation is given, pointing to the crystal-APD coupling.
\end{abstract}


\end{frontmatter}


\section{Introduction}
CALIFA (CALorimeter for In-Flight detection of gamma-rays and light charged pArticles) \cite{Cortina,califaSim}, is the calorimeter detector of the R$^{3}$B (Reactions with Relativistic Radioactive Beams) experiment at FAIR (Facility for Anti-proton and Ion Research), Darmstadt, Germany \cite{R3Bweb}. It consists of 2432 detection units of long CsI Tl-dopped scintillator crystals (refractive index at 1.79), with Large Area Avalanche Photo-Diode (LAAPD, or simply APD, refractive index at 1.55) based readout, arranged in two sections. The barrel section comprises 1952 detection units covering a polar angle from 43$^\mathrm{o}$ to 140$^\mathrm{o}$. The forward end-cap section has 480 detection units surrounding the target area, with polar angle coverage from 19$^\mathrm{o}$ to 43$^\mathrm{o}$. The system has full azimuthal coverage, a big detection dynamic range from around 100 keV gamma rays up to 320 MeV protons and an energy resolution below 6\% at 1 MeV after a gamma ray event reconstruction. Some modular fractions of the barrel section were constructed in recent years and extensively tested under different environments and conditions \cite{Pietras,Cabanelas}.

During the preparation of the CALIFA commissioning, a strong degradation in the behaviour of some detection units was observed as compared to their initial performance. In particular, the low energy noise increased by more than a factor of 10, the gain reduced by a factor of two and in some cases detection units even suffered some APD-crystal detachment, losing thus completely the optical coupling. Lack of light transmission can result in higher low energy noise and attenuation of photons causes loss of information and therefore can induce non-linearities in the detector response. Hence the optical properties of the material play a very important role in the overall efficiency of the system. The degraded detection units were initially included in the first modules of the CALIFA Demonstrator already evaluated and tested under realistic conditions in 2014 \cite{Pietras}. For the present study, detection units with that degradation are known as Group A.

On the other hand, such degradation was not observed neither in other detection units mounted around the same date, and tested in 2016 \cite{Cabanelas}, nor in other detection units mounted later and installed directly in the CALIFA Demonstrator modules for the commissioning. These detection units are known as Group B.

The detection units were extensively evaluated in the laboratory at the University of Santiago de Compostela in 2014 during their first assembly, and a new evaluation under the same conditions was performed late in 2018 to the degraded units of Group A in order to establish the causes of their loss of performance. In addition, some other detection units from Group B were re-evaluated as well. This paper addresses the analysis of the data collected with both groups of detection units, studies the possible causes of degradation and reports on the adopted solution.  

A more complete and exhaustive analysis of the properties and performance of the CALIFA CsI(Tl) scintillator crystals has recently been completed and can be found in \cite{lund}.

\section{Measurements description}
\begin{figure}[htb!]
\includegraphics[width=0.48\textwidth]{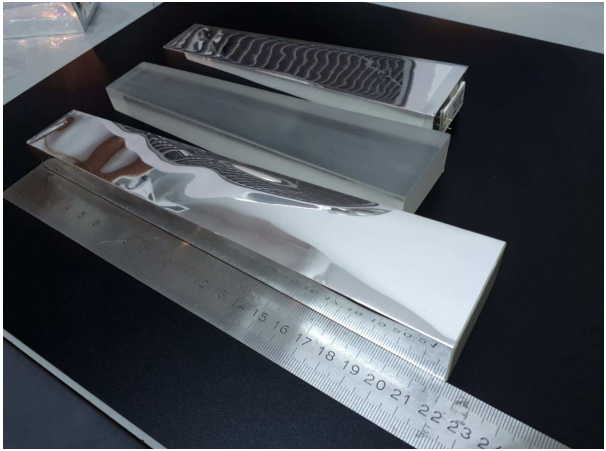}
\caption{Photograph of three crystals used for the measurements. From top to botton: 170 mm long, crystal with wrapping and coupled to the APD; 180 mm long, scintillator crystal only; 220 mm long, crystal with wrapping but without APD.} \label{fig_crystals_1}
\end{figure}

For the Group A, six different detection units from the CALIFA barrel detector, with different shapes and lengths, were evaluated after four years of operation. They consist of CsI(Tl) scintillator crystals from Amcrys$^{\mathrm{TM}}$, wrapped with 3M Vikuiti$^{\mathrm{TM}}$ reflective foil, and  coupled to Hamamatsu LAAPD model S8664-1020, commissioned as part of our industrial partnership. All crystals have a truncated pyramid shape with rectangular bases, as can be observed in Fig. \ref{fig_crystals_1}. The geometry of the crystals are 220 mm, 180 mm and 170 mm, in groups of two: crystals labelled as numbers 1 and 2 are 220 mm long, numbers 3 and 4 are 180 mm long, while crystals 5 and 6 are 170 mm long. The crystals in the calorimeter have different lengths due to the fact that forward focused protons at higher energy require longer crystals for full absorption.

Group B comprises another six detection units with the same geometrical and composition characteristics as those for Group A. Thus, crystals labeled as numbers 7 and 8 are 220 mm long, numbers 9 and 10 are 180 mm long, while crystals 11 and 12 are 170 mm long. The only difference between units from both groups are the first crystal-APD coupling date and the optical glue used.  

A dedicated setup in our laboratory was used to perform the measurements. Radioactive sources of $^{207}$Bi, $^{137}$Cs and $^{60}$Co were positioned in increments along the longitudinal crystal axis guided by a stepping motor, and the gamma emission was pin hole collimated with a lead filter. The different gamma ray spectra were recorded in steps of 20 mm. Thus, eight points were measured for the four shorter crystals (170 mm and 180 mm) while ten points were measured for the longer ones (220 mm). The APD read-out was done with a Mesytec MPRB-32 pre-amplifier \cite{Mesytec} and data was acquired with an ORTEC Trump multichannel analyzer PCI-Bus plug-in card \cite{Ortec}. The APDs were operated at their nominal voltages, corresponding to a gain of $M$=50, and small corrections to the HV as a function of the room temperature were applied on-the-fly as needed to compensate for the sensor dependence on the temperature \cite{lund}. The stability of the setup was studied and no variations in collected data were observed throughout the time.

For each detection unit, the relative energy resolution and the non-uniformity in Light Output, $\Delta$LO, was measured. The $\Delta$LO is the variation in the amount of light collected by the photosensor. It affects the energy resolution, and depends on the longitudinal position at which most of the energy deposition occurs in the crystal. The Light Output, LO, meanwhile, depends on several factors like the crystal shape, wrapping, surface polishing, transparency and optical coupling between the sensor and the crystal. Also, the dopant-concentration gradient can affect the $\Delta$LO and LO in long crystals. Thus, for a fixed-energy ionizing particle, the photopeak position in the ADC is a direct measurement of the LO. Therefore, by measuring a given photopeak position as a function of the distance to the photosensor, we are able to obtain the variations on the LO in the detection unit.

\begin{figure}[htb!]
\includegraphics[width=0.48\textwidth]{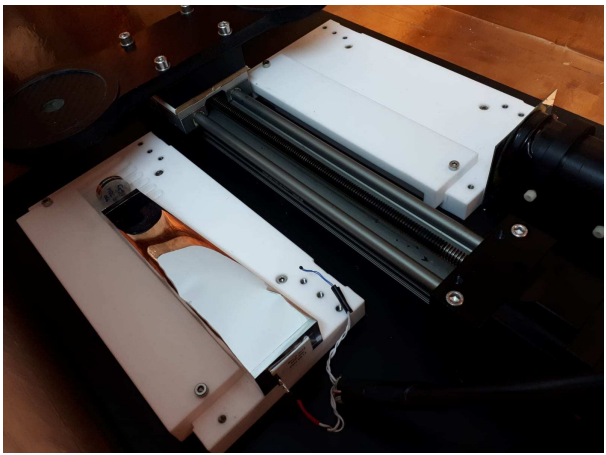}
\caption{View of one of the crystals inside the measurement setup together with the stepping motor for driving the radioactive sources. A radioactive source is seen, placed for a frontal irradiation measurement.} \label{fig_crystals_2}
\end{figure}

Fig. \ref{fig_crystals_2} shows one of the crystals, with the wrapping and the APD, ready to be evaluated in the setup where the stepping motor (upper part) can move the radioactive source and a lead collimator along the crystal longitudinal axis.

Five different sets of measurements, in different conditions, were performed to each crystal in Group A, named from Set A1 to Set A5. The readout was done with the originally associated APD for each crystal, except in one of the measurements, where the same reference APD was used for all crystals. Different optical adhesives have been used to couple the APD to the crystal in the different measurement sets. The precise compounds used for the optical coupling are named in the list of the different sets. The crystal wrapping was the same material for all sets (3M Vikuiti$^{\mathrm{TM}}$), but rearranged when removing APDs. Then, the different sets of measurements have the following characteristics: 

\begin{enumerate}
\item Set A1: In spring 2014, the six evaluated detection units were assembled and measured for the first time. Their corresponding APDs were coupled to the crystals with optical cement Scionix RTV 861 (refractive index at 1.43) \cite{Scionix}, an RTV based silicon two component glue for scintillators, PMTs and light guides.
\item Set A2: After four years of operation installed in a CALIFA crystal assembly, in 2018 they were removed from the mechanical structure and evaluated again.
\item Set A3: Later in 2018 the original APD and the optical cement were removed from the crystal and the detection units were then re-evaluated with the same original APD but the optical coupling was performed with the Rhodorsil P\^{a}te 7 optical grease (refractive index at 1.41) \cite{Rhod7}.
\item Set A4: In order to crosscheck also the response of the original APDs, all crystals were evaluated attached to a common reference APD. The optical coupling was made with Rhodorsil P\^{a}te 7 optical grease.
\item Set A5: Finally, the detection units were assembled again with their original APDs but the coupling was undertaken using optical colorless bi-component epoxy cement Eljen Technology EJ500 (refractive index at 1.57) \cite{EJ500}, suitable for optically bonding scintillators and acrylic materials.
\end{enumerate}

For Group B, only the three sets of measurements listed below were performed to detection units, labelled from B1 to B3:

\begin{enumerate}
\item Set B1: The six evaluated detection units for this group were assembled and measured for the first time in summer 2014 and their corresponding APDs were coupled to the crystals with optical mono-component cement Elec. Mic. Sci. Meltmount$^{\mathrm{TM}}$ 1.704 (refractive index at 1.70) \cite{MM17}, specifically formulated for use in microscope slide mounting and in other optical coupling applications.
\item Set B2: In spring 2019 they were removed from the mechanical structure where they were operating and evaluated again.
\item Set B3: This third set of measurements were only applied to the three crystals from Group B which showed worst performance from Set B2 with respect to Set B1. Those detection units were disassembled and then re-assembled with the optical cement EJ500.
\end{enumerate}

Both optical cements used in Group B have very similar optical properties and light transmission capabilities. Although it is a mono-component cement, Meltmount$^{\mathrm{TM}}$ 1.704 is not easy to manipulate and is sensitive to temperature and removable at about 60$^{o}$C; in addition, a expiration date of two years is given by the manufacturer. On the contrary, EJ500 has a better manipulation and results in a clearly harder union. Due to this last property, a further APD removal is not feasible when glued with EJ500. 

Table \ref{tab_sets} summarises the main characteristics of the two groups of sets of measurements previously presented.

\begin{table}[htb!]
\begin{tabular}{c c c c}
\hline
Set A & Year & Coupling adhesive & APD \\
\hline
A1 & 2014 & Scionix RTV861 & Original \\
A2 & 2018 & Scionix RTV861 & Original \\
A3 & 2018 & Rhodorsil P7 grease & Original \\
A4 & 2018 & Rhodorsil P7 grease & Reference \\
A5 & 2018 & Eljen T. EJ500 & Original \\
\hline
Set B & Year & Coupling adhesive & APD \\
\hline
B1 & 2014 & Melmount 1.7 & Original \\
B2 & 2019 & Melmount 1.7 & Original \\
B3 & 2019 & Eljen T. EJ500 & Original \\
\hline 
\end{tabular}
\caption{Summary of the main characteristics for the two groups of sets of measurements used in this work.} \label{tab_sets}
\end{table}

\section{Methods, Results and Data analysis}

\begin{figure}[htb!]
\includegraphics[width=0.49\textwidth]{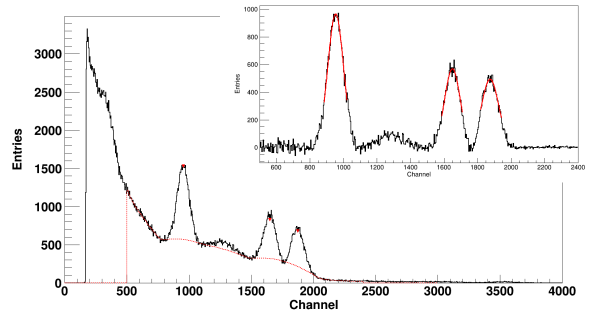}
\caption{Example of a measured $\gamma$-ray spectrum for $^{137}$Cs and $^{60}$Co radioactive sources and the calculated background using the ROOT ``ShowBackground'' estimation function (red dashed line), together with the Gaussian fit (insert, red solid line) over the photopeaks. The insert displays a zoom of the photopeaks after background subtraction.} \label{fig_spectrum}
\end{figure}

Each $\gamma$-ray spectrum was analyzed and the mean $M$ and the sigma $\sigma$ of photopeaks were extracted after a background subtraction and an automatic two-step recursive Gaussian fit. Figure \ref{fig_spectrum} shows an example with $\gamma$-ray detection from $^{137}$Cs and $^{60}$Co sources.

\begin{figure}[htb!]
\includegraphics[width=0.48\textwidth]{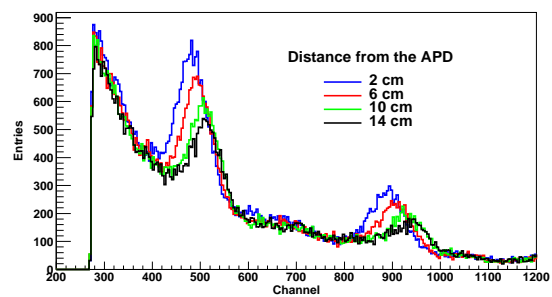}
\caption{Measured spectra for collimated $^{207}$Bi radioactive source at different positions along one of the 180 mm long crystals from Group A, coupled to APD with Scionix RTV 861. The two gamma rays from the source (0.569 MeV and 1.064 MeV) were detected. The picture illustrates the shifting of the photopeaks as a function of the distance of the source from the APD.} \label{fig_spectrum_distances}
\end{figure}

Figure \ref{fig_spectrum_distances} illustrates the effect of the $\Delta$LO. Four different spectra from the collimated $^{207}$Bi source over different positions along the same crystal are presented. For a given gamma ray, the collected light in the photosensor is different depending on the interaction position in the crystal, and accordingly its location in the spectrum, changes.

The relative energy resolution for a given energy photopeak is then calculated by:

\begin{center}
\begin{equation}
Res = \frac{\Delta E}{E} \times 100\% = \frac{FWHM}{E} \times 100\%
\label{eq_eres}
\end{equation}
\end{center}

being $FWHM$ the full width at half maximum, corresponding to $FWHM=2.35\sigma$ for a Gaussian fit. The average of the resolution for all positions photopeaks along the crystal is taken as the crystal resolution for the given energy. Such assumption was asserted by the fact that the obtained value is fully compatible with the single one obtained for traditional frontal irradiation.

The final value of energy resolution is calculated by fitting the corresponding energy resolution values of all measured photopeaks to a function $Res(E) = \frac{a}{\sqrt{E}} + b$ \cite{knoll}, widely confirmed to be the best match with experimental data in this kind of crystals \cite{Pietras2}, and evaluating it at $E$=1 MeV. The detected photopeaks energies are the 0.569 MeV and 1.064 MeV from $^{207}$Bi, the 1.173 MeV and 1.332 MeV from $^{60}$Co, and the 0.662 MeV from $^{137}$Cs.

The $\Delta$LO can be defined as \cite{Gong,Beylin}:

\begin{center}
\begin{equation}
\Delta LO = \frac{M_{Max}-M_{Min}}{M_{Avg}} \times 100\%
\label{eq_lonu}
\end{equation}
\end{center}

where $M_{Max}$ and $M_{Min}$ are the maximum and minimum mean of the fitted photopeak, in channels, for the set of all positions measured over a crystal, and $M_{Avg}$ is their average. The Normalized Light Output, NLO, can be defined for each of the measured positions as \cite{Ren}:

\begin{center}
\begin{equation}
NLO_{i} = \frac{M_{i}}{M_{Avg}}
\label{eq_nlo}
\end{equation}
\end{center}

where $M_{i}$ is the mean of the corresponding fitted photopeak measured at position $i$ over the crystal. Then, the preferable result corresponds to a value of NLO=1.

\begin{figure}[htb!]
\includegraphics[width=0.48\textwidth]{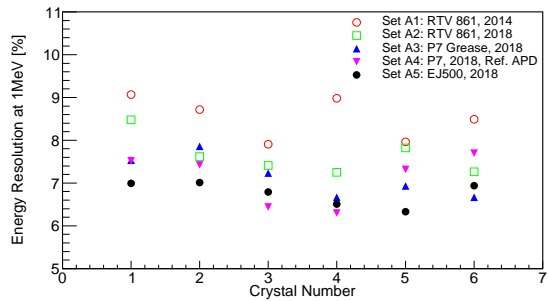}
\caption{Energy resolution at 1 MeV for the six detection units in Group A. The worst resolution (open red circles) was observed in Set A1, where the APDs were coupled to the crystals with RTV 861 optical cement, while a significant improvement is reached for Set A5 (full circles). Energy resolution is maintained for Set A5 at the level of 7\%.} \label{fig_RES}
\end{figure}

The energy resolution, $\Delta$LO and NLO values were calculated for all crystals in this evaluation. The energy resolution evaluated at 1 MeV for Group A is presented in Figure \ref{fig_RES}. The worst scenario corresponds to the first assembling data set, when RTV 861 optical cement was used. Sets with optical grease coupling improve the energy resolution in the same relative amount in all detection units, and no special contribution is observed from different APDs, which makes apparent that the main contribution comes from the scintillator crystal itself. Clearly, the EJ500 optical cement coupling of the crystal and the APD resulted in a better energy resolution for all crystals as compared with the results obtained with RTV 861, reaching now the acceptance criterion of 7\% in FWHM at 1 MeV for individual detection units. Also, low-energy noise events were reduced with all units showed approximately the same gain. There was little correlation between the crystal geometries and the energy resolution in this study, despite the fact that, in general for these geometries, the shorter the crystal, the better the energy resolution \cite{gascon}.

\begin{figure}[htb!]
\includegraphics[width=0.48\textwidth]{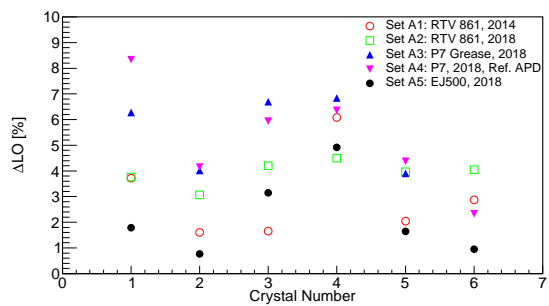}
\caption{Non-uniformity in Light Output in Group A. Sets with optical grease show a highest $\Delta$LO (both upright and flipped full triangles), while the lowest was reached with the EJ500 optical cement (full black circles).} \label{fig_LONU}
\end{figure}

The $\Delta$LO of each crystal in the five sets of data of Group A is presented in Figure \ref{fig_LONU}. Similar to the energy resolution, a significant improvement in the $\Delta$LO is reached in Set A5, after the final re-glueing with EJ500 optical cement, up to a factor of 4 in some cases. On the other hand, sets with the optical coupling done with Rhodorsil P\^{a}te 7 grease showed a far inferior $\Delta$LO behaviour in all cases. That effect highlights that the optical grease is an easy, clean and quick solution for initial bench tests, but it is not the most effective solution for long term operation detection units.

\begin{figure}[htb!]
\includegraphics[width=0.48\textwidth]{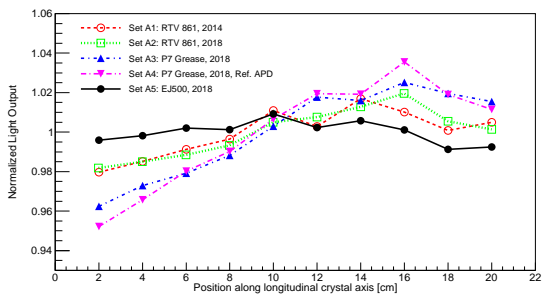}
\caption{Normalized Light Output for a 22-cm crystal in Group A as a function of the distance from the illumination spot to the APD face. It is observed the significant improvement of the $\Delta$LO along the crystal after re-glueing the APD with EJ500 optical cement (full black circles), and also how the worst measurements correspond to the optical grease cases (full triangles).} \label{fig_normLO}
\end{figure}

Figure \ref{fig_normLO} shows the NLO in one of the longer crystals of Group A (crystal number 1), measured with the 1.173 MeV $\gamma$ ray from the $^{60}$Co source. The $x$-axis represents the longitudinal distance in cm from the APD. It is clear that measurements with the EJ500 optical cement gave the most stable NLO around the value of 1. On the other hand, the NLO was the most uneven when the optical grease was used (Set A3 and Set A4). Since no fundamental reason has been determined for that result yet, this can point to the fact that higher $\Delta$LO contributions are strongly motivated by a lack of light transmission between crystal and APD and not only by crystal properties like dopant concentration or internal reflections. In the specific case of the EJ500 cement, two main factors are devised for its better light transmission as compared with other composites. First, the hardness and reliability of the APD-crystal union performed with that cement. Second, the almost perfect match between the refractive indexes of APD and EJ500 might play a crucial role ensuring a good light transmission. On the other hand, NLO was generally found to exceed 1 as the source is placed closer to the APD face of the crystal. That behaviour was already observed and studied in long CsI scintillation crystals in Ref. \cite{Ren}.

Table \ref{tab_results1} summarizes the more significant values calculated in this work for crystals of Group A. The statistical errors of the methods are computed and presented in the table as well.

\begin{table*}[htb!]
\centering
\begin{tabular}{|c|c|c|c|c|c|c|c|}
\hline
Crystal & Length & E. Res. [\%] & E. Res. [\%] & E. Res. [\%] & $\Delta$LO [\%] & $\Delta$LO [\%] & $\Delta$LO [\%] \\
number & [mm] & RTV 861 & Rhod. P7 & EJ500 & RTV 861 & Rhod. P7 & EJ500 \\ 
 & & Set A1 & Set A3 & Set A5 & Set A1 & Set A3 & Set A5 \\
\hline
1 & 220 & 9.1 $\pm$ 0.3 & 7.5 $\pm$ 0.9 & 7.0 $\pm$ 0.5 & 3.7 $\pm$ 0.2 & 6.3 $\pm$ 0.3 & 1.8 $\pm$ 0.1 \\
2 & 220 & 8.7 $\pm$ 0.5 & 7.9 $\pm$ 0.7 & 7.0 $\pm$ 0.7 & 1.1 $\pm$ 0.1 & 4.0 $\pm$ 0.2 & 0.8 $\pm$ 0.1 \\
3 & 180 & 7.9 $\pm$ 0.2 & 7.2 $\pm$ 0.6 & 6.8 $\pm$ 0.4 & 1.6 $\pm$ 0.1 & 6.7 $\pm$ 0.3 & 3.1 $\pm$ 0.2 \\
4 & 180 & 9.0 $\pm$ 0.5 & 6.7 $\pm$ 0.5 & 6.5 $\pm$ 0.7 & 6.1 $\pm$ 0.3 & 6.8 $\pm$ 0.2 & 5.0 $\pm$ 0.3 \\
5 & 170 & 8.0 $\pm$ 0.4 & 6.9 $\pm$ 0.7 & 6.3 $\pm$ 0.3 & 2.0 $\pm$ 0.1 & 3.9 $\pm$ 0.2 & 1.6 $\pm$ 0.2 \\
6 & 170 & 8.5 $\pm$ 0.6 & 7.7 $\pm$ 1.0 & 6.9 $\pm$ 0.3 & 2.9 $\pm$ 0.1 & 2.3 $\pm$ 0.1 & 0.9 $\pm$ 0.1 \\
\hline
\end{tabular}
\caption{Summary of energy resolution and $\Delta$LO measurements for Group A detection units. Three sets of measurements are shown in the table: Set A1, year 2014, optical coupling with cement RTV 861; Set A3, year 2018, optical coupling with grease Rhodorsil P7; Set A5, year 2018, optical coupling with cement EJ500. It is clear how the detection units improved their performance after re-glueing. On the other hand, crystal geometry does not play a strong role in the units characteristics. See text for a complete discussion. Computed statistical errors are shown.} \label{tab_results1}
\end{table*}

Together with the improvements in light output non-linearity and energy resolution, the problem of the higher low energy noise observed in some degraded units was solved after recovery as well. Since both the crystals and APDs remained in good conditions, this confirmed the suspicions that the lost of optical coupling resulted also in bigger noise contributions. It is worth to mention that keeping a good light collection efficiency and transmission ensures that our measurements are only limited by the intrinsic noise contribution of the LAAPDs.  

\begin{figure}[htb!]
\includegraphics[width=0.48\textwidth]{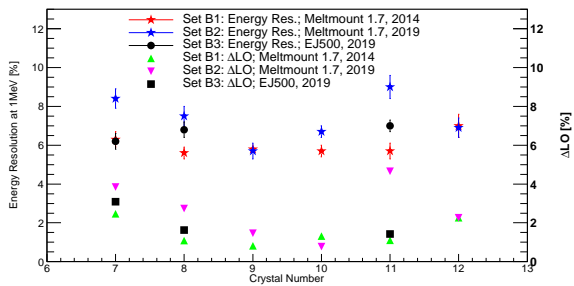}
\caption{Energy Resolution and $\Delta$LO for the three sets of measured performed to the crystals of Group B. Both $\Delta$LO and energy resolution improves significantly after re-glueing the APD with EJ500 optical cement in Set B3 (open circles and crosses). In addition, it is noticeable also the degradation observed in some of the evaluated units of Set B2 (units 7, 8 and 11) with respect to their performances in Set B1, which indeed led us to take the decision of re-glueing those units.} \label{fig_groupb}
\end{figure}

Concerning results of Group B, although the behaviour observed for those detection units assembled with optical glue Meltmount$^{\mathrm{TM}}$ 1.7 was good in general, still in some cases a small degradation occurred. Given the fact that the mentioned optical glue is easily removable, a sample of the Melmount 1.7 coupled units were disassembled and re-coupled using optical glue EJ500. The improvements in $\Delta$LO and energy resolution motivated the decision to convert the remainder to EJ500. Such observation is clear in Figure \ref{fig_groupb}: it is shown both the energy resolution and the $\Delta$LO calculated for the six units of Group B in the three different sets of measurements. Detection units numbering 7, 8 and 11 showed a degradation close to 30\% in energy resolution after five years assembled with Meltmount$^{\mathrm{TM}}$ 1.7, exceeding thus the acceptance limit of 7\%. But after re-assembling with EJ500, in Set B3 a significant improvement was achieved in energy resolution and also in $\Delta$LO, and the units went back within the requirements. For data of Group B as seen in Figure \ref{fig_groupb}, computed statistical errors are at the order of 5\% for the $\Delta$LO, and not higher than 7\% for the energy resolution.

Once again, it seems evident that the lost of performance from detection units in mid/long term operation is directly related with the quality of the crystal-APD junction, that is, which optical compound is used for the coupling. Moreover, since manipulations when assembling were done following all the manufacturer's specifications, we believe that the intrinsic lifetime of the used compounds plays the more important role in degradation, together with how accurate are them for the precise junction of an APD with the CsI polished face. Then, we can claim that no appreciable aging effects can be attributed neither to the CsI(Tl) crystals nor the Avalanche Photo Diodes. It is also advisable to keep as small as possible the difference between the respective refractive indices of crystals, optical glue and APDs in order to reduce internal reflection in light transmission. Long-term performance of EJ500 in this context remains to be evaluated, but it is expected to be superior for reasons predefined. On the other hand, it is known that crystals can suffer from radiation damage during in-beam operation, leadinf to the formation of color centres reducing light transmission, which results in a decrease of the photon yield and therefore in a light uniformity deterioration \cite{Dementyev, Ren}. However, changes in light yield and light output degradations related with radiation damage are not expected due to the short running periods and low radiation level of the R$^{3}$B experiments \cite{R3Bweb}. In addition, it was evaluated the effect of radiation on LAAPDs \cite{califaTDR} and the conclusion is that radiation damage is not a detrimental factor over the lifetime of the APDs either.  

As a summary, Table \ref{tab_results2} encloses the more significant conclusions and observations for the four different analyzed adhesives.

\begin{table*}[htb!]
\centering
\begin{tabular}{|c|c|c|c|c|}
\hline
Coupling Adhesive & Type & Removable & Observations \\
\hline
Scionix RTV861 \cite{Scionix} & Cement & No (expected) & Degradation with time; unreliable junction \\
Rhodorsil P7 \cite{Rhod7} & Grease & Yes & Lower light transmission capabilities \\
Eljen Tech. EJ500 \cite{EJ500} & Cement & No & Overall best performances \\
Melmount 1.7 \cite{MM17} & Cement & Yes (at high T) & Hard to manipulate; expiration date \\
\hline
\end{tabular}
\caption{Summary of the main observations in this work for the four different coupling adhesives used. As stated in the text, we recommend the use of the Eljen Technology EJ500 adhesive to be used as a final coupling compound between APDs and CsI scintillator crystals.} \label{tab_results2}
\end{table*}

\section{Conclusions}

Two groups of 6 detection units of the CALIFA calorimeter, each consisting of CsI scintillating crystals with APD readout, of the R$^{3}$B experiment at FAIR, have been investigated in terms of optical coupling, energy resolution and non-uniformity in Light Output following mid/long term operation. The crystals were irradiated with $\gamma$-ray radioactive sources in a dedicated bench setup.

After a period of four and five years of operation, keeping the units in a controlled humidity environment, no changes or degradation were observed in the scintillator crystal properties. Also, the performance of the APDs remained unaltered. We expect that all CALIFA detection units can be operated for long time periods without performance degradation or aging effects. However, degradation from long-term high-energy particle exposure remain to be evaluated.

After recovering the units doing the coupling between the crystal and the APD with the most appropriate optical cement, they improved significantly in performance: the energy resolution remained below 7\% (the acceptance criterion for individual units) improving by up to 20\% with respect to the initial conditions. For the $\Delta$LO, an improvement of a factor of two was observed in most cases and the NLO reduced, approaching the expected value of 1 along the full longitudinal dimension of the crystal. Then, the observed performance degradation can be attributed to an optical cement issue, either related with manipulation during assembly or due to the lack of stability and lifetime of the applied compound, being the latter the most important contribution. Thus, an accurate choice in the optical coupling material is essential for the long term stability of the detection units. Optical cement Eljen Technology EJ500 was found to be the optimal choice for our purposes.

When coupling crystal and APD with optical grease, the $\Delta$LO of the detection units was systematically inferior by a factor between 0.3 and 4, depending on the unit, with respect to optical cements coupling, while no significant loss of energy resolution was observed in that case. No significant differences were observed between original associated and reference APDs. We suggest that light transmission effects play a very important role in the $\Delta$LO measurements.

\section*{Acknowledgements}

This work has been financially supported by the European Union Horizon 2020 research and innovation programme under grants agreements No 262010 (ENSAR) and No 654002 (ENSAR2), the Spanish MICCIN grants FPA47831-C2-1P and FPA2015-69640-C2-1-P, by the Plan Galego de Investigaci\'on, Innovaci\'on e Crecemento (I2C) of Xunta de Galicia under projects POS-B/2016/015, GRC2013-011 and ED431C 2017/54 and by the German BMBF (No. 05P19RDFN1), TU Darmstadt - GSI cooperation contract, HIC for FAIR. 

\section*{References}

\bibliography{mybibfile}

\end{document}